\magnification 1190
\baselineskip 15 pt

\def \P {{\bf P}}
\def \E {{\bf E}}
\def \F {{\bf F}}
\def \C {{\cal C}}

\def \L {{\cal L}}
\def \R {{\bf R}}

\def \la {\lambda}
\def \al {\alpha}

\def \sig {\sigma}
\def \td {\tau_d}
\def \to {\tau_0}
\def \noi {\noindent}

\def \hphi {\hat \phi}
\def \ws {w_s}

\def \zo {z_0}
\def \zb {z_b}
\def \zbo {z_{b0}}

\input srctex.sty

\def \sect#1{\bigskip  \noindent{\bf #1} \medskip }
\def \subsect#1{\bigskip \noindent{\it #1} \medskip}
\def \thm#1#2{\medskip \noindent {\bf Theorem #1.}   \it #2 \rm \medskip}
\def \prop#1#2{\medskip \noindent {\bf Proposition #1.}   \it #2 \rm \medskip}
\def \cor#1#2{\medskip \noindent {\bf Corollary #1.}   \it #2 \rm \medskip}
\def \pf {\noindent  {\it Proof}.\quad }
\def \lem#1#2{\medskip \noindent {\bf Lemma #1.}   \it #2 \rm \medskip}

\def \rem#1{\medskip \noindent {\bf Remark #1.}}

\def\sqr#1#2{{\vcenter{\vbox{\hrule height.#2pt\hbox{\vrule width.#2pt height#1pt \kern#1pt\vrule width.#2pt}\hrule height.#2pt}}}}

\def \square{\hfill\mathchoice\sqr56\sqr56\sqr{4.1}5\sqr{3.5}5}

\def \qed {$\square$ \medskip}

\nopagenumbers

\headline={\ifnum\pageno=1 \hfill \else \hfill {\rm \folio} \fi}

\centerline{\bf Optimally Investing to Reach a Bequest Goal}

\bigskip

\centerline{Erhan Bayraktar and Virginia R. Young}
\centerline{Department of Mathematics, University of Michigan}
\centerline{Ann Arbor, Michigan, USA, 48109} \bigskip

\centerline{Version:  20 May 2016} \bigskip

\noindent{\bf Abstract:}  We determine the optimal strategy for investing in a Black-Scholes market in order to maximize the probability that wealth at death meets a bequest goal $b$, a type of goal-seeking problem, as pioneered by Dubins and Savage (1965, 1976).  The individual consumes at a constant rate $c$, so the level of wealth required for risklessly meeting consumption equals $c/r$, in which $r$ is the rate of return of the riskless asset.  

Our problem is related to, but different from, the goal-reaching problems of Browne (1997).  First, Browne (1997, Section 3.1) maximizes the probability that wealth reaches $b < c/r$ before it reaches $a < b$.  Browne's game ends when wealth reaches $b$.  By contrast, for the problem we consider, the game continues until the individual dies or until wealth reaches 0; reaching $b$ and then falling below it before death does not count.

Second, Browne (1997, Section 4.2) maximizes the expected discounted reward of reaching $b > c/r$ before wealth reaches $c/r$.  If one interprets his discount rate as a hazard rate, then our two problems are {\it mathematically} equivalent for the special case for which $b > c/r$, with ruin level $c/r$.  However, we obtain different results because we set the ruin level at 0, thereby allowing the game to continue when wealth falls below $c/r$.

%  However, if one identifies Browne's discount rate in his Section 4.2 with the constant hazard rate in this work, then the problems are identical.  That said, Browne considers only the case for which wealth is greater than the level required to fund consumption risklessly, and we shall see that our solution is new because we choose our ruin level to be strictly less than this safe level for consumption, unlike in Browne's work.  Furthermore, we also solve the problem when wealth is less than the safe level for consumption and when the bequest goal itself is less than this safe level.

 %  But, the more interesting case is when these two problems are not the same. %We, thereby, make more objective the goal of maximizing expected utility at death, first considered in a continuous-time framework by Merton (1969).  Specifically, instead of requiring the individual to choose a utility function, we only require the individual to choose a bequest goal $b$.  We learn that, for wealth lying between $0$ and $b$, the optimal investment strategy is {\it independent} of $b$, a surprising result.  Therefore, if the individual were to revise her bequest goal, her investment strategy would not change if her wealth is less than the new goal.

\medskip


\medskip


\sect{1. Introduction}

We determine the optimal strategy for investing in a Black-Scholes market in order to maximize the probability that wealth at death meets a bequest goal $b$, a problem considered in part by Browne (1997, Section 4.2).  We, thereby, make more objective the goal of maximizing expected utility of death, first considered in a continuous-time framework by Merton (1969).  Specifically, instead of requiring the individual to choose a utility function, we only require the individual to choose a bequest goal $b$.  We learn that, for wealth lying between $0$ and $b$, the optimal investment strategy is {\it independent} of $b$, a surprising result.  Therefore, if the individual were to revise her bequest goal, her investment strategy would not change if her wealth is less than the new goal.

Our paper falls naturally within the area of optimally controlling wealth to reach a goal.  Research on this topic began with the seminal work of Dubins and Savage (1965, 1976) and continued with the work of Pestien and Sudderth (1985), Orey et al.\ (1987), Sudderth and Weerasinghe (1989), Kulldorff (1993), Karatzas (1997), and Browne (1997, 1999a, 1999b).   A typical problem considered in this research is to control a process to maximize the probability the process reaches $b$, either before a fixed time $T$, such as in Karatzas (1997), or before the process reaches $a < b$, such as in Pestien and Sudderth (1985).  In either of these forms of the problem, the game ends if wealth reaches $b$.  The problem we consider in this paper is similar in that we control a wealth process to maximize the probability of reaching $b$ before $0$, but we want to reach $b$ at a random time, namely, the time of death of the investor.  The game does not end if wealth reaches $b$ before the investor dies; the game only ends when the individual dies or ruins.

Our problem is related to, but different from, the goal-reaching problems of Browne (1997).  First, Browne (1997, Section 3.1) maximizes the probability that wealth reaches $b < c/r$ before it reaches $a < b$.  Browne's game ends when wealth reaches $b$.  By contrast, for the problem we consider, the game continues until the individual dies or until wealth reaches 0.  Second, Browne (1997, Section 4.2) maximizes the discounted reward of achieving a goal $b \ge c/r$ if $W_ 0 \in [c/r, b]$; if one interprets his discount rate as a hazard rate, then our two problems are {\it mathematically} equivalent for the special case for which $b \ge c/r$, with ruin level $c/r$.  ($W_t$ is the individual's wealth at time $t \ge 0$, $c$ is the constant rate of consumption, and $r$ is the rate of return on the riskless asset.  Thus, $c/r$ is the amount of wealth required to fund consumption risklessly.)  However, Browne's solution (1997, Section 4.2) implicitly restricts investment strategies to be such that if $W_0 \in [c/r, b]$, then $W_t \in [c/r, b]$ almost surely, for all $t \ge 0$.  By contrast, we do not restrict our investment strategies in this manner.  Furthermore, we solve the bequest problem when initial wealth $W_0 = w < c/r \le b$ and when the bequest goal $b < c/r$.

The rest of the paper is organized as follows. In Section 2, we present the financial market in which the individual invests, we formalize the problem of maximizing the probability of reaching a bequest goal, and we give a verification lemma that will help us to find that maximum probability, along with the optimal strategy for investing in the financial market.  In Section 3, we solve the problem of maximizing the probability of reaching a bequest goal when the rate of consumption is $0$; we separate this case because we can solve it explicitly.  Sections 4 and 5 parallel Section 3 for a positive rate of consumption.  When the rate of consumption is positive, we cannot obtain the maximum probability of reaching the bequest goal explicitly, but we can solve the problem for its convex Legendre dual, and we do so in Sections 4 and 5, specifically, in Section 4.2 and 5.2.

For the case considered in Section 4, the convex Legendre dual is the value function for an optimal stopping problem, which we present in Section 4.1.  For the case considered in Section 5, the convex Legendre dual is the solution of a time-homogeneous, two-phase Stefan problem, and we present the corresponding free-boundary problem in Section 5.1.  In Sections 3.2, 4.3, and 5.2, we study properties of the optimal investment strategy and discover that the optimal amount to invest in the risky asset is {\it independent} of the bequest goal $b$ when wealth is less than $b$, a surprising result.  Sections 6 and 7 conclude the paper; in those sections, we compare our work with that of Browne (1997) and summarize our results, respectively.

\sect{2.  Statement of the problem and verification lemma}

In this section, we present the financial market for the investor.  Then, we state the optimization problem this investor faces and present a verification lemma that we will use to solve the optimization problem.

\subsect{2.1. Financial market and probability of reaching the bequest goal}

We assume the individual has an investment account she manages in order to reach a given bequest goal $b > 0$.  She consumes from this account at the constant rate $c \ge 0$.  The individual invests in a Black-Scholes financial market with one riskless asset earning interest at the rate $r > 0$ and one risky asset whose price process $S = \{ S_t \}_{t \ge 0}$ follows geometric Brownian motion:
$$
dS_t = \mu \, S_t \, dt + \sigma \, S_t \, dB_t,
$$
in which $B = \{ B_t \}_{t \ge 0}$ is a standard Brownian motion on a filtered probability space $(\Omega, {\cal F}, \F = \{ {\cal F}_t \}_{t \ge 0}, \P)$, with $\mu > r$ and $\sigma > 0$.

Let $W_t$ denote the wealth in the individual's investment account at time $t \ge 0$.  Let $\pi_t$ denote the dollar amount invested in the risky asset at time $t \ge 0$.  An investment policy $\Pi = \{ \pi_t \}_{t \ge 0}$ is {\it admissible} if it is an $\F$-progressively measurable process satisfying $\int_0^t \pi^2_s \, ds < \infty$ almost surely, for all $t \ge 0$.  Thus, wealth follows the dynamics
$$
\left\{
\eqalign{
dW_t &= (r W_t + (\mu - r) \pi_t - c) dt + \sigma \, \pi_t \, dB_t, \cr
W_0 &= w \ge 0.
}
\right.
\eqno(2.1)
$$

Denote the future lifetime random variable of the investor by $\td$; suppose $\td$ follows an exponential distribution with mean $1/\la$.   We assume the individual seeks to maximize the probability that $W_{\td} \ge b$, by optimizing over admissible controls $\Pi$.  We do not insist admissible strategies be such that $W_t \ge 0$ almost surely, for all $t \ge 0$, because of the constant drain on wealth by the negative drift term $- c$ when $c > 0$. Therefore, we end the game if wealth reaches $0$ before the individual dies.  Define $\to = \inf \{ t \ge 0: W_t \le 0 \}$, and define the value function by
$$
\phi(w) = \sup_{\Pi} \P^{w} \left( W_{\td \wedge \to} \ge b \right),
\eqno(2.2)
$$
in which $\P^{w}$ denotes conditional probability given $W_0 = w \ge 0$.

\rem{2.1} {If wealth is large enough, say, at least $\ws$ (``s'' for safe), then the individual can invest all her wealth in the riskless asset with the interest income sufficient to cover her consumption and with $W_{\td} \ge b$ almost surely.  This so-called {\it safe level} is given by
$$
\ws = \max \left( b, \, {c \over r} \right).
\eqno(2.3)
$$
Thus, $\phi(w) = 1$ if $w \ge w_s$, and it remains for us to determine $\phi(w)$ for $0 < w < \ws$.}

\subsect{2.2 Verification lemma}

In this section, we provide a verification lemma that states that a classical solution of a boundary-value problem (BVP) associated with the maximization problem in (2.2) equals the maximum probability of reaching the bequest goal.  Therefore, we can reduce our problem to one of solving a BVP.  We state the verification lemma without proof because its proof is similar to others in the literature; see, for example, Bayraktar and Young (2007).  First, for $\pi \in \R$, define a differential operator $\L^\pi$ by its action on a test function $f$.
$$
\eqalign{
\L^\pi \, f &= (rw + (\mu - r) \pi - c) f_w + {1 \over 2} \sigma^2 \pi^2 f_{ww} - \la \left(f - {\bf 1}_{\{ w \ge b \}} \right).
}
\eqno(2.4)
$$

\lem{2.1} {Let $\Phi = \Phi(w)$ be a $\C^2$ function that is non-decreasing and concave on $[0, \ws]$, except perhaps at $b$, where it will be $\C^1$ and have left- and right-second derivatives.  Suppose $\Phi$ satisfies the following boundary-value problem.
$$
\left\{
\eqalign{
&\max_\pi \L^\pi \, \Phi(w) = 0, \cr
&\Phi(0) = 0, \quad \Phi(\ws) = 1.
}
\right.
\eqno(2.5)
$$
\hfill \break
\noi Then, on $[0, \ws]$,
$$
\phi = \Phi,
$$
and the optimal amount invested in the risky asset is given in feedback form by
$$
\pi^*_t = - {\mu - r \over \sig^2} \, {\phi_w(W^*_t) \over \phi_{ww}(W^*_t)},
\eqno(2.6)
$$
for all $t \in [0, \td \wedge \to)$, in which $W^*_t$ is optimally controlled wealth at time $t$.  \qed}

We use Lemma 2.1 to calculate $\phi$.  The solution differs depending on whether $c = 0$, $0 < c \le rb$, or $c > rb$, so we split the problem into those three cases in the next three sections, respectively.  Specifically, in Section 3, we consider the case for which $c = 0$ and explicitly determine $\phi$.  In Sections 4 and 5, we consider the two cases for which $c > 0$ and express $\phi$ through its Legendre dual.

\sect{3. The case for which $c = 0$}

In Section 3.1, we obtain an explicit expression for the maximum probability of reaching the bequest goal and the corresponding optimal investment strategy.  In Section 3.2, we study properties of that optimal investment strategy.

\subsect{3.1 Maximum probability of reaching the bequest goal}

When $c = 0$, the safe level $\ws$ equals $b$.  From Lemma 2.1, we know if we find an increasing, concave solution of the following BVP on $[0, b]$, then that solution equals the maximum probability of reaching the bequest goal.  By slightly abusing notation, we write $\phi$ in the statement of the BVP.
$$
\left\{
\eqalign{
& \la \phi = rw \phi_w + \max_\pi \left[(\mu - r) \pi \phi_w + {1 \over 2} \sig^2 \pi^2 \phi_{ww} \right], \cr
& \phi(0) = 0,  \quad \phi(b) = 1.
}
\right.
\eqno(3.1)
$$

We give the solution of this BVP in the next theorem, along with the optimal investment strategy in the risky asset.  We omit the proof because it is a straightforward application of Lemma 2.1.

\thm{3.1} {If $c = 0$, the maximum probability of reaching the bequest goal equals
$$
\phi(w) = \left( {w \over b} \right)^q,  \qquad 0 \le w \le b,
\eqno(3.2)
$$
in which
$$
q = {1 \over 2r} \left[ (r + \la + m) - \sqrt{(r + \la + m)^2 - 4 r \la} \right] \in (0, 1),
\eqno(3.3)
$$
and
$$
m = {1 \over 2} \left( {\mu - r \over \sig} \right)^2.
$$
When wealth equals $w \in (0, b)$, the optimal amount invested in the risky asset is given by
$$
\pi^*(w) = {\mu - r \over \sig^2} \, {w \over 1 - q}.
\eqno(3.4)
$$}

\rem{3.1} {Browne (1997, Section 4.2) maximizes the expected value of the discounted hitting time $\tau_b$ of $b$ when $c/r \le w \le b$, that is, he maximizes $\E^w \left[ e^{-\la \tau_b} \right]$ with the understanding that the game ends if wealth reaches $c/r$.  If we interpret the discount rate as the hazard rate, then his problem is identical to ours when $c = 0$.  With this correspondence, we observe that the optimal amount to invest in the risky asset given in Theorem 4.2 of Browne (1997) when $c = 0$ equals the expression in (3.4), as one would expect. \qed}

\rem{3.2} {Note that $\phi$ in (3.2) decreases as bequest goal $b$ increases, which is expected from the definition of $\phi$.  Even though ruin is impossible under the optimal strategy, the investor might die with wealth less than $b$, and as $b$ increases, the probability of reaching the bequest goal decreases.  Furthermore, because $q$ in (3.3) increases with $\la$, the probability of reaching the bequest goal decreases with $\la$, which is intuitively pleasing.  Indeed, as the individual becomes more likely to die sooner rather than later, reaching the bequest goal becomes less likely.  Finally, because $q$ decreases with $m$, the probability of reaching the bequest goal increases with $m$. This result makes sense because as the return on the risky market becomes more favorable--either from larger drift $\mu$ or from lower volatility $\sig$--the probability of reaching the bequest goal increases.  \qed}

\rem{3.3} {We find it notable that the optimal investment strategy in (3.4) is {\it independent} of the bequest goal $b$ when wealth is less than $b$.  Also, the investment strategy is {\it identical} to the one employed by an investor who maximizes the expected discounted utility of her wealth at death under the utility function $u(w) = w^q$, that is, with constant relative risk aversion of $1 - q \in (0, 1)$, in which $q$ is given in (3.3).  Specifically, the maximum-utility problem is $\sup_{\Pi} \E^w \left[ e^{-\rho \td} \, (W_{\td})^q \right]$, for some $\rho > 0$.  Thus, if we were to observe an individual investing a constant proportion of her wealth in a risky asset, then we could say she is maximizing the expected discounted utility of her wealth at some time in the future or maximizing the probability her wealth at death equals a specific bequest goal.  This correspondence is similar to the one found by Bayraktar and Young (2007), in which they relate the optimal strategies for maximizing the individual's expected utility of lifetime consumption and for minimizing her probability of lifetime ruin. \qed}

\cor{3.2} {If $c = 0$, then $W^*$, the optimally controlled wealth process, follows the dynamics
$$
d W^*_t = W^*_t  \left[ \left( r + {2m \over 1 - q} \right) dt + {\mu - r \over \sig} \, {1 \over 1- q} \, dB_t \right], \quad 0 < W^*_t < b.
$$
Thus, $W^*_t > 0$ almost surely, for all $t \ge 0$, if $W_0 = w \in (0, b)$.  \qed}

\rem{3.4} {Corollary 3.2 states that because $W^*$ follows geometric Brownian motion, ruin will not occur when investing optimally.  Thus, the maximum probability of reaching the bequest goal equals the probability $W^*$ hits the safe level $b$ before the individual dies, which equals a particular value of the Laplace transform of the corresponding hitting time.  Indeed, when $c = 0$, $\phi(w) = \E^w \left(e^{-\la \tau_b} \right)$, in which $\tau_b = \inf \{ t \ge 0: W^*_t \ge b \}$. \qed}

\subsect{3.2 Properties of the optimal investment strategy}

In this section, we present two corollaries of Theorem 3.1 in which we explore properties of the optimal investment strategy given in (3.4).  As the individual becomes more likely to die sooner rather than later, we expect her to invest more in the risky asset to reach her goal before dying.  By contrast, as the risky asset becomes more volatile, the individual does not need to invest as much wealth in the risky asset to reach her bequest goal.  These intuitive expectations are confirmed by the results of Corollaries 3.3 and 3.4 below.

First, we determine when the investment strategy results in leveraging, that is, when the individual borrows from the riskless asset to invest more than her current wealth in the risky.  Note that, from the expression for the optimal investment strategy given in (3.4), we deduce $\pi^*(w) > w$ for any $w > 0$ if and only if
$$
{\mu - r \over \sig^2} \, {1 \over 1 - q} > 1.
\eqno(3.5)
$$

\cor{3.3} {If $c = 0$, then leveraging occurs at all levels of wealth between $0$ and $b$ either if $\la \ge {\mu + r \over 2}$ or if $\la < {\mu + r \over 2}$ and $\sig < \sig_l$ for some $\sig_l > 0$.}

\pf One can show that $\sig^2 (1 - q)$ increases with respect to $\sig$; thus, the left side of inequality (3.5), ${\mu - r \over \sig^2} \, {1 \over 1 - q}$, decreases with respect to $\sig$.  As $\sig$ approaches 0, $\sig^2 (1 - q)$ also approaches 0, so ${\mu - r \over \sig^2} \, {1 \over 1 - q}$ approaches $\infty$.

Also, as $\sig$ approaches $\infty$, the left side of (3.5) converges to
$$
\lim_{\sig \rightarrow \infty} {\mu - r \over \sig^2} \, {1 \over 1 - q} = 
\cases{0, &if $r \ge \la$, \cr \cr
{2(\la - r) \over \mu - r}, &if $r < \la$,}
$$
and the second expression is less than 1 if and only if $\la < {\mu + r \over 2}$.  The statements in the corollary follow from these observations.  \qed

From the observation in the proof of Corollary 3.3 that $\sig^2 (1 - q)$ increases with respect to $\sig$ and from the fact that $q$ increases with respect to $\la$, as noted in Remark 3.2, we obtain the following corollary.

\cor{3.4} {If $c = 0$, then the optimal amount invested in the risky asset increases with respect to $\la$ and decreases with respect to $\sig$. \qed}

\sect{4. The case for which $0 < c \le rb$}

When the rate of consumption is less than $rb$, the safe level $\ws$ equals $b$, as in Section 3.  However, when the rate of consumption is positive, then we cannot write $\phi$ explicitly, as in Theorem 3.1.  In Section 4.1, we introduce an auxiliary optimal stopping problem; then, in Section 4.2, we show that its concave Legendre transform is equal to the maximum probability of reaching the bequest goal.  Finally, in Section 4.3, we study properties of the optimal investment strategy.

\subsect{4.1 A related optimal stopping problem}

Consider the following payoff function $u$ defined for $z \in \R^+$.
$$
u(z) = \max(1 - bz, 0) = (1 - bz)_+.
\eqno(4.1)
$$
%Fix values $0 \le \zb < {1 \over b} < \zo$. Note that $u$ is minimal among those convex functions $f$ defined on $\R^+$ that satisfy
%$$
%f(\zb) = 1 - b \zb, \hbox{ and } f_z(\zb) = -b,
%$$ 
%and
%$$
%f(\zo) = 0 = f_z(\zo).
%$$
Define a stochastic process $Z = \{ Z_t \}_{t \ge 0}$ by
$$
dZ_t = (\la - r) Z_t \, dt + {\mu - r \over \sig} \, Z_t \, d \hat B_t,
\eqno(4.2)
$$
in which $\hat B = \{ \hat B_t \}_{t \ge 0}$ is a standard Brownian motion on a filtered probability space $(\hat \Omega, \hat {\cal F}, \hat \F = \{\hat {\cal F}_t \}_{t \ge 0}, \hat \P)$, and consider the optimal stopping problem given by
$$
\hphi(z) = \sup_\tau \hat \E^z \left[ - \int_0^\tau c \, e^{-\la t} \, Z_t \, dt + e^{-\la \tau} (1 - b Z_\tau)_+ \right],
\eqno(4.3)
$$
in which the supremum is taken over stopping times with respect to $(\hat \Omega, \hat {\cal F}, \hat \F, \hat \P)$.

The game embodied in (4.3) charges the player the running cost $c Z_t$ between now and the time he stops, discounted by the survival probability $e^{-\la t}$.  At the time of stopping $\tau$, he receives $u(Z_\tau) = (1 - b Z_\tau)_+$.  Thus, the player has to decide whether it is better to continue playing the game by paying $c Z_t$ continually or to stop and take the payoff $(1 - b Z_\tau)_+$.

Note that $\hphi$ is convex.  Indeed, because $Z_t = z H_t$, with
$$
H_t = \exp \left( -(r - \la + m) t + {\mu - r \over \sig} \hat B_t \right),
$$
we can write the integral in (4.3) as
$$
- z \, \int_0^\tau c \, e^{-\la t} \, H_t \, dt,
$$
a linear function of $z$.  In addition, $u$ is convex, so the expectation is convex.  Finally, because the supremum of convex functions is convex, $\hphi$ is convex.  Similarly, because both the integral and $u$ are non-increasing, it follows that $\hphi$ is non-increasing.

Define the {\it continuation region} by
$$
C = \{ z \in \R^+: \hphi(z) > (1- bz)_+ \},
$$
so the optimal time to stop is $\tau^* = \inf \{ t \ge 0: Z_t \not \in C \}$.  By following the line of argument in Section 2.7 of Karatzas and Shreve (1998), we can assert that there exist $0 \le \zb < {1 \over b} < \zo$ (to be determined) such that $C = (\zb, \zo)$.  Thus, we can rewrite the optimal time to stop as $\tau^* = \inf \{ t \ge 0: Z_t \le \zb \hbox{ or } Z_t \ge \zo \}$.  Furthermore, this argument shows that $\hphi$ is the unique classical solution of the following free-boundary problem (FBP) on $[\zb, \zo]$.
$$
\left\{
\eqalign{
& \la \hphi = (\la - r) z \hphi_z + m z^2 \hphi_{zz} - cz, \cr
& \hphi(\zb) = 1 - b \zb, \quad \hphi_z(\zb) = -b,  \cr
& \hphi(\zo) = 0 = \hphi_z(\zo).
}
\right.
\eqno(4.4)
$$
For $0 \le z < \zb < {1 \over b}$, $\hphi(z) = (1 - b z)_+ = 1 - bz$ because it is optimal to stop when $z$ is less than $\zb$.  For $z > \zo > {1 \over b}$, $\hphi(z) = (1 - bz)_+ = 0$ because it is optimal to stop when $z$ is greater than $\zo$.  Also see Peskir and Shiryaev (2006) for results concerning the equivalence of optimal stopping problems and FBPs.

In the following proposition, we present the solution of the FBP (4.4).

\prop{4.1} {Suppose $0 < c \le rb$.  The solution of the free-boundary problem $(4.4)$ on $[\zb, \zo]$ and, hence, the value function of the optimal stopping problem  $(4.3)$, is given by
$$
\hphi(z) = {c \over r} \, \zo \, \left[ {1 - \al_2 \over \al_1 - \al_2} \left( {z \over \zo} \right)^{\al_1} +  {\al_1 - 1 \over \al_1 - \al_2} \left( {z \over \zo} \right)^{\al_2} - {z \over \zo} \right],
\eqno(4.5)
$$
in which
$$
\eqalign{
\al_1 &= {1 \over 2m} \left[ (r - \la + m) + \sqrt{(r - \la + m)^2 + 4 m \la} \right] > 1, \cr
\al_2 &= {1 \over 2m} \left[ (r - \la + m) - \sqrt{(r - \la + m)^2 + 4 m \la} \right] < 0.
}
\eqno(4.6)
$$
The free boundary $\zo > {1 \over b}$ is given by
$$
\zo = {\zb \over \zbo},
$$
in which $\zbo \in (0, 1)$ uniquely solves
$$
{c \over r} \left[{\al_1 (1 - \al_2) \over \al_1 - \al_2} \, \zbo^{\al_1 - 1} + {\al_2 (\al_1 - 1) \over \al_1 - \al_2} \, \zbo^{\al_2 - 1} \right] =  {c \over r} - b,
\eqno(4.7)
$$
and the free boundary $\zb < {1 \over b}$ is given in terms of $\zbo$ by
$$
{1 \over \zb} = {c \over r} \, {(\al_1 - 1)(1 - \al_2) \over \al_1 - \al_2} \left( - \zbo^{\al_1 - 1} +  \zbo^{\al_2 - 1} \right).
\eqno(4.8)
$$
Moreover, $\hphi$ is $\C^2$ and is decreasing and convex on $[\zb, \zo]$.}

\pf First, note that there exists a unique solution $\zbo \in (0, 1)$ of (4.7).  Indeed, the left side of (4.7) increases with respect to $\zbo$; as $\zbo$ approaches $0+$, the left side of (4.7) approaches $-\infty$; and, when $\zbo = 1$, the left side equals ${c \over r} > {c \over r} - b$.  

It is easy to show that the expression in (4.5) satisfies the differential equation in (4.4) and that it satisfies the free-boundary conditions $\hphi(\zo) = 0 = \hphi_z(\zo)$.  The expressions in (4.7) and (4.8) imply $\hphi$ in (4.5) satisfies the free-boundary conditions $\hphi(\zb) = 1 - b \zb$ and $\hphi_z(\zb) = -b$.

We wish to show that $\zb < {1 \over b} < \zo$.  From (4.7) and (4.8), we see that the inequality $\zb < {1 \over b}$ holds if and only if 
$$
 {1 - \al_2 \over \al_1 - \al_2} \, \zbo^{\al_1 - 1} + {\al_1 - 1 \over \al_1 - \al_2} \, \zbo^{\al_2 - 1} > 1.
 \eqno(4.9)
$$
The left side of (4.9) decreases with respect to $\zbo$ on $(0, 1]$ and equals 1 when $\zbo = 1$; thus, (4.9) holds for all $\zbo \in (0, 1)$, from which it follows that $\zb < {1 \over b}$.  Similarly, $\zo > {1 \over b}$ if and only if ${\zbo \over \zb} < b$, or equivalently via (4.8),
$$
1 + {(\al_1 - 1)(1 - \al_2) \over \al_1 - \al_2} \left( \zbo^{\al_1} -  \zbo^{\al_2} \right) - {\al_1 (1 - \al_2) \over \al_1 - \al_2} \, \zbo^{\al_1 - 1} - {\al_2 (\al_1 - 1) \over \al_1 - \al_2} \, \zbo^{\al_2 - 1} > 0.
\eqno(4.10)
$$
It is straightforward to show that the left side of inequality (4.10) decreases with respect to $\zbo$ on $(0, 1]$ and equals $0$ when $\zbo = 1$; thus, (4.10) holds for all $\zbo \in (0, 1)$, from which it follow that $\zo > {1 \over b}$.

Finally, we show that $\hphi$ given in (4.5) is, indeed, decreasing and convex on $[\zb, \zo]$, as expected, because $\hphi$ defined in (4.3) uniquely solves (4.4).  To that end, observe that
$$
\hphi_z(z) = {c \over r} \, \left[ {\al_1(1 - \al_2) \over \al_1 - \al_2} \left( {z \over \zo} \right)^{\al_1-1} +  {\al_2(\al_1 - 1) \over \al_1 - \al_2} \left( {z \over \zo} \right)^{\al_2-1} - 1 \right],
$$
and
$$
\hphi_{zz}(z) = {c \over r} \, {(\al_1 - 1)(1 - \al_2) \over \al_1 - \al_2} \, \left[ \al_1 \left( {z \over \zo} \right)^{\al_1-2} - \al_2 \left( {z \over \zo} \right)^{\al_2-2}\right] > 0.
$$
Because $\hphi_{zz}(z) > 0$ and $\hphi_z(\zo) = 0$, we conclude that $\hphi$ given in (4.5) is decreasing and convex on $[\zb, \zo]$. \qed

In the next section, we show that the solution of the FBP (4.4) is intimately connected with the maximum probability of reaching the bequest goal.

\hangindent 16 pt \subsect{4.2 Relation between the optimal stopping problem and the maximum probability of reaching the bequest goal}

In this section, we show that the Legendre transform (see, for example, Karatzas and Shreve (1998)) of the solution of the FBP (4.4) is, in fact, the maximum probability of reaching the bequest goal. To this end, note that because $\hphi$ in (4.3) or (4.5) is convex, we can define its concave dual via the Legendre transform.

\prop{4.2} {Suppose $0 < c \le rb$.  Define $\Phi$ on $[0, b]$ by
$$
\Phi(w) = \min_{\zb \le z \le \zo} \left[ \hphi(z) + wz \right],
\eqno(4.11)
$$
in which $\hphi$ is the value function of the optimal stopping problem in $(4.3)$. Then, the maximum probability of reaching the bequest goal equals $\Phi$ on $[0, b]$.}

\pf The optimizer $z^*$ of (4.11) solves the equation $\hphi_z(z) + w = 0$; thus, $z^* = I(-w)$, in which $I$ is the functional inverse of $\hphi_z$.  Recall $\hphi_z < 0$ on $(\zb, \zo)$.  It follows that
$$
\Phi(w) = \hphi(I(-w)) + w I(-w).
$$
This expression implies that $\Phi_w(w) = I(-w)$; thus, $z^* = \Phi_w(w)$.  Moreover, $\Phi_w(w) = I(-w)$ implies that $\Phi_{ww}(w) = -1/\hphi_{zz}(I(-w))$.  It follows that $\Phi$ is increasing and concave on $[0, b]$.

By using these relationships and by substituting $z = I(-w) = \Phi_w(w)$ into $\hphi$'s FBP (4.4), we deduce that $\Phi$ solves the following BVP.
$$
\left\{
\eqalign{
& \la \Phi = (rw - c) \Phi_w + \max_\pi \left[(\mu - r) \pi \Phi_w + {1 \over 2} \sig^2 \pi^2 \Phi_{ww} \right], \cr
& \Phi(0) = 0,  \quad \Phi(b) = 1.
}
\right.
\eqno(4.12)
$$
In (4.12), we use $\zo = \Phi_w(0)$ and $\zb = \Phi_w(b)$, which follow from the free-boundary definitions of $\zo$ and $\zb$.  It follows from Lemma 2.1 that $\phi = \Phi$ on $[0, b]$.  \qed

We combine the results of Propositions 4.1 and 4.2 in the following theorem.

\thm{4.3} {If $0 < c \le rb$, then the maximum probability of reaching the bequest goal is given by
$$
\phi(w) = {c \over r} \, {(\al_1 - 1)(1 - \al_2) \over \al_1 - \al_2} \left[- \left( {z \over \zo} \right)^{\al_1 - 1} + \left( {z \over \zo} \right)^{\al_2 - 1} \right] z,
\eqno(4.13)
$$
in which $\zo$ is given in Proposition $4.1$.  Here, for a given $w \in [0, b]$, $z \in [\zb, \zo]$ uniquely solves
$$
{c \over r} \left[{\al_1(1 - \al_2) \over \al_1 - \al_2} \, \left( {z \over \zo} \right)^{\al_1 - 1} + {\al_2(\al_1 - 1) \over \al_1 - \al_2} \, \left( {z \over \zo} \right)^{\al_2 - 1} \right] = {c \over r} - w.
\eqno(4.14)
$$
When wealth equals $w$, the optimal amount invested in the risky asset is given by
$$
\pi^*(w) = {\mu - r \over \sig^2} \, {c \over r} \, {(\al_1 - 1)(1 - \al_2) \over \al_1 - \al_2} \left[ \al_1 \left( {z \over \zo} \right)^{\al_1 - 1} - \al_2 \left( {z \over \zo} \right)^{\al_2 - 1} \right].
\eqno(4.15)
$$}

\rem{4.1} {We remind the reader that the game ends if the investor's wealth reaches $0$ before she dies.  By contrast, as mentioned in Remark 3.1, for wealth lying between $c/r$ and $b$, Browne (1997, Section 4.2) effectively maximizes the probability of reaching the bequest goal $b$ before reaching $c/r$, if we interpret his parameter $\la$ as a hazard rate.  In the proof of his Theorem 4.2, Browne chose a solution that forced his value function to be $0$ at $c/r$.  Thus, he tacitly imposed the condition that the game ends if wealth reached $c/r$ before dying. (Alternatively, he implicitly restricted admissible investment strategies to be such that $W_t \ge c/r$ almost surely for all $t \ge 0$ if $W_0 = w > c/r$.)  Because our ruin level of $0$ is less than Browne's, except when $c = 0$, our maximum probability of reaching the bequest goal (before ruin) is larger than the expression he found in his Theorem 4.2 for $c/r \le w \le b$.  In the next section, we compare the optimal investment strategy in (4.15) with the one in Browne's Theorem 4.2. \qed}

%Before examining properties of the optimal investment strategy given in (4.15), we present the following corollary of Theorem 4.3 describing how the maximum probability of reaching the bequest goal changes with some of the underlying parameters.

%\cor{4.5} {If $0 < c \le rb$, then the maximum probability of reaching the bequest goal satisfies the following properties.
%\item{$(i)$} $\phi$ decreases as the bequest goal $b$ increases.
%\item{$(ii)$} $\phi$ decreases as the rate of consumption $c$ increases.
%\item{$(iii)$} $\phi$ decreases as the hazard rate $\la$ increases.
%\item{$(iv)$} $\phi$ increases as $m$ increases.}

%\pf

%\qed

\subsect{4.3 Properties of the optimal investment strategy}

The first, and most surprising, result is that the optimal amount to invest in the risky asset is {\it independent} of the bequest goal $b$ when wealth is less than $b$.  We observed this in the case for which $c = 0$, but it is also true when $0 < c \le rb$.

\prop{4.4} {Consider two bequest goals, $b_1 < b_2$.  If $0 < c \le rb_1$, then the optimal amounts to invest in the risky asset under the two bequest goals are identical when wealth is less than $b_1$.}

\pf The proof is simple.  The value of ${z \over \zo}$ that solves (4.14) is independent of $b$; thus, $\pi^*(w)$ in (4.15) is independent of $b$.  \qed

\rem{4.2} {Young (2004) found a similar result when minimizing the probability of lifetime ruin with a fixed rate of consumption.  The optimal amount to invest in the risky asset was independent of the ruin level for wealth greater than the ruin level.  Thus, if the investor's preferred ruin level were to change, her investment strategy would not.  Similarly, for our problem of maximizing the probability of reaching a bequest goal, if the investor's preferred bequest goal were to change, her investment strategy would not. \qed}

We next determine when the optimal investment strategy in Theorem 4.3 is increasing or decreasing with respect to wealth.

\prop{4.5} {If $0 < c \le rb$, then the following statements indicate how $\pi^*$ varies with respect to wealth:
\item{$(i)$} If $r \le \la$, then $\pi^*$ is increasing on $[0, b]$.
\item{$(ii)$} If $\la < r < \la + m$, then $\pi^*$ is decreasing on $[0, w^*)$ and increasing on $(w^*, b]$, for some $w^* \in (0, b)$.
\item{$(iii)$} If $r \ge \la + m$ and if $0 < c < c^*$, for some $c^* \in (0, \, rb)$, then $\pi^*$ is decreasing on $[0, w^*)$ and increasing on $(w^*, b]$, for some $w^* \in (0, b)$.
\item{$(iv)$} If $r \ge \la + m$ and if $c \ge c^*$, then $\pi^*$ is decreasing on $[0, b]$.}

\pf  By differentiating the expression for $\pi^*(w)$ in (4.15) with respect to $w$, we obtain
$$
{d \pi^*(w) \over d w} \propto \left[ \al_1(\al_1 - 1) \left( {z \over \zo} \right)^{\al_1 - 1} + \al_2(1 - \al_2) \left( {z \over \zo} \right)^{\al_2 - 1}  \right] {\partial \over \partial w} \left( {z \over \zo} \right).
$$
Then, by differentiating (4.14) fully with respect to $w$, we learn
$$
\left[\al_1 \left( {z \over \zo} \right)^{\al_1 - 2} - \al_2 \left( {z \over \zo} \right)^{\al_2 - 2} \right] {\partial \over \partial w} \left( {z \over \zo} \right) \propto - 1.
$$
Thus, because the expression in the square brackets is positive, we deduce that $z/\zo$ decreases with $w$, so
$$
{d \pi^*(w) \over d w} \propto - \left[ \al_1(\al_1 - 1) \left( {z \over \zo} \right)^{\al_1 - 1} + \al_2(1 - \al_2) \left( {z \over \zo} \right)^{\al_2 - 1}  \right] =: f(z).
\eqno(4.16)
$$
It is easy to see that $f$ decreases with $z \in [\zb, \zo]$.  Thus, $\pi^*(w)$ increases on all of $[0, b]$, or $f$ is positive on $[\zb, \zo)$, if and only if $f(\zo) \ge 0$, which is equivalent to $r \le \la$.

For the remainder of the proof, assume $\la < r$; then, $f(\zo) < 0$ and $\pi^*(w)$ is decreasing at $w = 0$.  Because $f$ decreases on $[\zb, \zo]$, if $f(\zb) > 0$, then $\pi^*(w)$ first decreases and then increases on $[0, b]$.  Similarly, if $f(\zb) \le 0$, then $\pi^*(w)$ decreases on all of $[0, b]$.  So, we have reduced the proof to showing when $f(\zb) > 0$.

To that end, let $z = \zb$ in the expression for $f$ in (4.16), substitute for $\zbo^{\al_2 - 1}$ from (4.7), and simplify to obtain
$$
f(\zb) \propto \al_1 \, {-r + \la + m \over m} \,  \zbo^{\al_1 - 1} + (1 - \al_2) \left( {rb \over c} - 1 \right).
\eqno(4.17)
$$
Note that if $r < \la + m$, then $f(\zb) > 0$ automatically.  Thus, we consider the case for which $r \ge \la + m$; $f(\zb) > 0$ if and only if
$$
\zbo^{\al_1 - 1} < {1 - \al_2 \over \al_1(\al_1 + \al_2 - 2)} \left( {rb \over c} - 1 \right).
\eqno(4.18)
$$ 
Recall that the left side of (4.7) increases with respect to $\zbo$.  Thus, inequality (4.18) holds if, when we substitute the right side of (4.18) for $\zbo^{\al_1 - 1}$ into the left side of (4.7), the (new) left side is greater than the right.  That is, (4.18) holds if and only if the following inequality holds.
$$
\eqalign{
&{(1 - \al_2)^2 \over (\al_1 - \al_2)(\al_1 + \al_2 - 2)} \left( {rb \over c} - 1 \right) + {\al_2(\al_1 - 1) \over \al_1 - \al_2} \left[ {1 - \al_2 \over \al_1(\al_1 + \al_2 - 2)} \left( {rb \over c} - 1 \right) \right]^{-{1 - \al_2 \over \al_2 - 1}} \cr
&\qquad > - \left( {rb \over c} - 1 \right),}
$$
or equivalently,
$$
\left[ {1 \over \al_1 + \al_2 - 2} \left( {rb \over c} - 1 \right) \right]^{{\al_1 - \al_2 \over \al_1 - 1}} > - {\al_2 \over \al_1 - 1} \left( {1 - \al_2 \over \al_1} \right)^{-{1 - \al_2 \over \al_1 - 1}}.
\eqno(4.19)
$$
Define $c^* \in (0, rb)$ to be the value such that the left side of (4.19) equals the right side.  For $c < c^*$, inequality (4.19) will hold; on the other hand, for $c \ge c^*$, inequality (4.19) will not hold and we have $f(\zb) \le 0$.  \qed

\rem{4.3} {The investor in our problem really faces two problems.  First, she maximizes the probability that her wealth at death is at least equal to $b$.  Second, she wants to avoid ruin because she cannot continue playing the game if she ruins.  Thus, we expect the optimal investment strategy in Theorem 4.3 to be a blend of the one in Theorem 3.1 in which the investor can avoid ruin because $c = 0$ and the one for an investor who seeks to minimize the probability that she ruins before dying with no bequest goal (Young, 2004).  Recall that the former is ${\mu - r \over \sig^2} \, {w \over 1 - q}$, and the latter is ${\mu - r \over \sig^2} \, {c/r - w \over p - 1}$, in which ${1 \over p - 1} = \al_1 - 1$.  The former is an increasing function of wealth; the latter, decreasing.  \hfill \break
\indent If $\la \ge r$, then, because her mortality rate is relatively large, the investor worries more about reaching her bequest goal and less about ruin, which is borne out by the fact that the optimal investment strategy acts more like the one in Theorem 3.1, that is, it is increasing on all of $[0, b]$.  At the other extreme, if $\la < r - m$ and if $c$ is large enough, then ruin is more of a concern, so the optimal investment strategy acts more like the one for minimizing the probability of lifetime ruin, that is, it is decreasing on all of $[0, b]$.  \hfill \break
\indent  Between these two extremes, the optimal investment strategy first decreases and then increases with wealth.  Thus, for wealth close to $0$, the individual invests similarly to one who seeks to avoid ruin in that the optimal investment strategy decreases with wealth; and, for wealth close to the bequest goal, the individual invests similarly to one who seeks to reach a bequest goal without the threat of ruin in that the optimal investment strategy increases with wealth. \qed}

This last observation leads to the questions:   When wealth is close to $0$, how does $\pi^*(w)$ compare with the optimal investment strategy of one who seeks to avoid ruin, that is, ${\mu - r \over \sig^2} \, {c/r - w \over p - 1}$?  When wealth is close to $b$, how does $\pi^*(w)$ compare with the optimal investment strategy of one who seeks to reach a bequest goal with the threat of ruin, that is, ${\mu - r \over \sig^2} \, {w \over 1 - q}$?  We answer these questions in the next two propositions.

In the next proposition, we show that, for our problem, the optimal amount to invest in the risky asset is greater than if we were to minimize the probability of lifetime ruin with no bequest goal.  This makes sense because in trying to reach a bequest goal, the investor has to take on risk to increase wealth; she is not merely avoiding ruin.

\prop{4.6} {If $0 < c \le rb$, and if wealth lies between $0$ and $c/r$, then 
$$
\pi^*(w) > {\mu - r \over \sig^2} \, {c/r - w \over p - 1}.
\eqno(4.20)
$$}

\pf After substituting for $\pi^*(w)$ from (4.15), substituting for $c/r - w$ from (4.14), and simplifying, we find that (4.20) is equivalent to $-\al_2(1 - \al_2) > \al_2(\al_1 - 1)$, which is true because the left side is positive, and the right is negative.  \qed

\prop{4.7} {If $0 < c \le rb$, and if the solution $\zbo$ of $(4.7)$ is such that $\al_1 \zbo^{\al_1-1} > 1$, then for all $0 \le w \le b$,
$$
\pi^*(w) > {\mu - r \over \sig^2} \, {w \over 1 - q}.
\eqno(4.21)
$$
Otherwise, if $\al_1 \zbo^{\al_1-1} < 1$, which occurs if $b$ is large enough, then $(4.21)$ holds on $[0, w^*)$, and the following holds on $(w^*, b]$, for some $w^* \in (0, b)$.
$$
\pi^*(w) < {\mu - r \over \sig^2} \, {w \over 1 - q}.
\eqno(4.22)
$$}

\pf  After substituting for $\pi^*(w)$ from (4.15) and simplifying, we find that (4.21) is equivalent to
$$
\al_1 \left( {z \over \zo} \right)^{\al_1 - 1} > 1.
\eqno(4.23)
$$
This inequality holds at $z = \zo$ because $\al_1 > 1$, which we expect because $\pi^*(0) > 0$.  The left side of (4.23) increases with $z$; thus, (4.21) holds on for all $0 \le w \le b$ if and only if (4.23) holds when $z = \zb$.

On the other hand, if $\al_1 \zbo^{\al_1-1} < 1$, which one can show holds if $b$ is large enough, then $\pi^*(b) < {\mu - r \over \sig^2} \, {b \over 1 - q}$, and (4.22) holds for wealth close enough to $b$. \qed

In the next proposition, we show that our problem is continuous with respect to $c$ as $c$ approaches $0$.

\prop{4.8} {As $c$ approaches $0$, $\phi$ and $\pi^*$ in Theorem $4.3$ approach $\phi$ and $\pi^*$, respectively, in Theorem $3.1$.}

\pf As $c$ approaches $0$, $\zbo$ approaches $0$, as does ${c \over r} \, \zbo^{\al_1 - 1}$.  From (4.7), it follows that ${c \over r} \, \zbo^{\al_2 - 1}$ approaches $- {\al_1 - \al_2 \over \al_2(\al_1 - 1)} \, b$. The two free boundaries $\zb$ and $\zo$ approach ${q \over b}$ and $\infty$, respectively, and from (4.14), it follows that ${c \over r} \left( {z \over \zo} \right)^{\al_2 - 1}$ approaches $- {\al_1 - \al_2 \over \al_2(\al_1 - 1)} \, w$, thereby generalizing the result when $w = b$.  Thus, $\left( {z \over \zb} \right)^{\al_2 - 1}$ approaches ${w \over b}$, from which it follows that $z$, the solution of (4.14) approaches ${q \over b} \left( {w \over b} \right)^{q - 1}$, in which we use the fact that $1 - q = {1 \over 1 - \al_2}$.

From these results, we deduce the following limit for $\phi$.
$$
\lim_{c \rightarrow 0} \phi(w) = {(\al_1 - 1)(1 - \al_2) \over \al_1 - \al_2} \, \lim_{c \rightarrow 0} {c \over r} \, \left( {z \over \zo} \right)^{\al_2 - 1} z = \left( {w \over b} \right)^q,
$$
which equals the probability of reaching the bequest goal when $c = 0$; see the expression in (3.2) in Theorem 3.1.  Similarly, $\pi^*$ has the following limit.
$$
\lim_{c \rightarrow 0} \pi^*(w) = {\mu - r \over \sig^2} \, {(\al_1 - 1)(1 - \al_2) \over \al_1 - \al_2} (- \al_2)  \lim_{c \rightarrow 0} {c \over r} \, \left( {z \over \zo} \right)^{\al_2 - 1} = {\mu - r \over \sig^2} \, {w \over 1 - q},
$$
which equals the optimal amount to invest in the risky asset when $c = 0$; see the expression (3.4) in Theorem 3.1.  \qed

In the next proposition, we compare $\pi^*$ with the optimal investment strategy in Browne (1997, Theorem 4.2).

\prop{4.9} {If $0 < c \le rb$, and if wealth lies between $c/r$ and $b$, then
$$
\pi^*(w) > {\mu - r \over \sig^2} \, {w - c/r \over 1 - q}
\eqno(4.24)
$$}

\pf  After substituting for $\pi^*(w)$ from (4.15), substituting for $w - c/r$ from (4.14), and simplifying, we learn that inequality (4.24) holds is equivalent to $\al_1 - \al_2 > 0$, which is true.  \qed

\rem{4.4} {${\mu - r \over \sig^2} \, {w - c/r \over 1 - q}$ is the optimal amount to invest if one is maximizing the probability of reaching the bequest goal $b$, with a ``ruin level'' of $c/r$, as in Browne (1997, Theorem 4.2).  Because our ruin level $0$ is less than $c/r$, the investor can take on more risk in the financial market to achieve her bequest goal.  She does not need to worry that her wealth might fall to $c/r$; if it does, she can continue playing the game.  However, in Browne (1997, Theorem 4.2), the individual invests in such a way that her wealth avoids reaching $c/r$, just as the individual in our Theorem 3.1 invests in such a way that she will not ruin.  \qed}

When minimizing the probability of lifetime ruin, the optimal amount invested in the risky asset increases as $c$ increases (Young, 2004).  The next proposition tells us that the same is true for the optimal investment strategy when wealth is near $0$, which makes sense because the investor wants to avoid ruin so that she may continue investing to reach the bequest goal.

\prop{4.10} {If $0 < c \le rb$, then $\pi^*$ increases with respect to $c$ for wealth close to $0$.}

\pf By differentiating (4.14) with respect to $c$, we learn that
$$
{1 \over y} \, {\partial y \over \partial c} = {rw \over c^2} {\al_1 - \al_2 \over (\al_1 - 1)(1 - \al_2)} \, {1 \over \al_1 y^{\al_1 - 1} - \al_2 y^{\al_2 - 1}},
$$
in which $y = {z \over \zo} \in [\zbo, 1]$.  Thus,
$$
\eqalign{
{\partial \pi^* \over \partial c} &\propto {\partial \over \partial c} \left[ c \left(\al_1 y^{\al_1 - 1} - \al_2 y^{\al_2 - 1} \right) \right] \cr
&= \left(\al_1 y^{\al_1 - 1} - \al_2 y^{\al_2 - 1} \right) + c \left(\al_1(\al_1 - 1) y^{\al_1 - 1} + \al_2(1 - \al_2) y^{\al_2 - 1} \right) {1 \over y} {\partial y \over \partial c} \cr
&\propto  \left(\al_1 y^{\al_1 - 1} - \al_2 y^{\al_2 - 1} \right)^2 +  {rw \over c} \, (\al_1 - \al_2) \left[{\al_1 \over 1 - \al_2} \, y^{\al_1 - 1} + {\al_2 \over \al_1 - 1} \, y^{\al_2 - 1}\right] \cr
&= \left(\al_1 y^{\al_1 - 1} - \al_2 y^{\al_2 - 1} \right)^2 \cr
& \quad + \left[ 1 - {\al_1(1 - \al_2) \over \al_1 - \al_2} \, y^{\al_1 - 1} - {\al_2(\al_1 - 1) \over \al_1 - \al_2} \, y^{\al_2 - 1} \right] \cr 
& \qquad \times (\al_1 - \al_2) \left[{\al_1 \over 1 - \al_2} \, y^{\al_1 - 1} + {\al_2 \over \al_1 - 1} \, y^{\al_2 - 1}\right] \cr
& \propto {\al_1(\al_1 - 1) \over \al_1 - \al_2} \, y^{1 - \al_2} + {\al_2 (1 - \al_2) \over \al_1 - \al_2} \, y^{1 - \al_1} - \al_1 \al_2.
}
$$
If $w = 0$, then $y = 1$, and from the above calculation, it follows that
$$
{\partial \pi^* \over \partial c}\bigg|_{w = 0} \propto (\al_1 - 1)(1 - \al_2) > 0.
$$
Thus, in a neighborhood of $w = 0$, the optimal amount invested in the risky asset increases with the rate of consumption.  \qed

\sect{5. The case for which $c > rb$}

This case differs from the two in the preceding sections because the safe level ${c \over r}$ is greater than the bequest goal $b$.  Thus, if the individual dies when wealth is at least $b$ but less than ${c \over r}$, she will have reached her bequest goal.  In Section 5.1, we introduce an auxiliary free-boundary problem.  Then, in Section 5.2, we show that its concave Legendre transform is equal to the maximum probability of reaching the bequest goal, and we study properties of the optimal investment strategy.

\subsect{5.1 A related free-boundary problem}

Consider the following FBP on $[0, \zo]$, with $0 < \zb < \zo$ to be determined.
$$
\left\{
\eqalign{
& \la \hphi = (\la - r) z \hphi_z + m z^2 \hphi_{zz} - cz + \la {\bf 1}_{\{ z \le \zb \}}, \cr
& \hphi(0) = 1, \cr
& \hphi_z(\zb) = -b,  \cr
& \hphi(\zo) = 0 = \hphi_z(\zo).
}
\right.
\eqno(5.1)
$$
This FBP is a time-homogeneous, two-phase Stefan problem (Fasano and Primicero, 1997), with transition boundary $\zb$ lying between two domains, one which has an additional driving term of $\la$.

In the following proposition, we present the solution of the FBP (5.1). We omit the proof because it is similar to the one for Proposition 4.1.

\prop{5.1} {Suppose $c > rb$.  The solution of the free-boundary problem $(5.1)$ on $[0, \zo]$ is given by
$$
\hphi(z) =
\cases{ 
1 + \left( {c \over r} - b \right) {\zb \over \al_1} \left( {z \over \zb} \right)^{\al_1} - {c \over r} \, z, &if $0 \le z \le \zb$, \cr \cr
{c \over r} \, \zo \, \left[ {1 - \al_2 \over \al_1 - \al_2} \left( {z \over \zo} \right)^{\al_1} +  {\al_1 - 1 \over \al_1 - \al_2} \left( {z \over \zo} \right)^{\al_2} - {z \over \zo} \right], &if $\zb < z \le \zo$,
}
\eqno(5.2)
$$
in which $\al_1$ and $\al_2$ are as in $(4.6)$. The free boundary $\zo$ is given by
$$
{1 \over \zo} = {c \over r} \, {\al_1 - 1 \over \al_1} \, \zbo^{\al_2}
\eqno(5.3)
$$
in which $\zbo \in (0, 1)$ uniquely solves $(4.7),$ and $\zb = \zo \zbo$.  Moreover, $\hphi$ is decreasing and convex on $[0, \zo]$, and it is $\C^2$, except at $z = \zb$ where it is $\C^1$ with left- and right-second derivatives. \qed}

%{1 \over \zo} = {c \over r} \, \left[{1 - \al_2 \over \al_1 - \al_2} \, \zbo^{\al_1} + {\al_1 - 1 \over \al_1 - \al_2} \, \zbo^{\al_2} \right] - \left( {c \over r} - b \right) {1 \over \al_1} \, \zbo,

In the next section, we show that the solution of the FBP (5.1) is intimately connected with the maximum probability of reaching the bequest goal.

\hangindent 17 pt \subsect{5.2 Relation between the free-boundary problem and the maximum probability of reaching the bequest goal}

In this section, we show that the Legendre transform of the solution of the FBP (5.1) is, in fact, the maximum probability of reaching the bequest goal when $c > rb$. To this end, note that because $\hphi$ in (5.2) is convex, we can define its concave dual via the Legendre transform, as in Section 4.2.

\prop{5.2} {Suppose $c > rb$.  Define $\Phi$ on $[0, c/r]$ by
$$
\Phi(w) = \min_{0 \le z \le \zo} \left[ \hphi(z) + wz \right],
\eqno(5.4)
$$
in which $\hphi$ is given in $(5.2)$. Then, the maximum probability of reaching the bequest goal equals $\Phi$ on $[0, c/r]$.}

\pf  As in the proof of Proposition 4.2, we deduce that $\Phi$ is an increasing, concave function of $w$ and solves the following BVP on $[0, c/r]$.
$$
\left\{
\eqalign{
& \la \left(\Phi - {\bf 1}_{\{ w \ge b \}} \right) = (rw - c) \Phi_w +  \max_\pi \left[(\mu - r) \pi \Phi_w + {1 \over 2} \sig^2 \pi^2 \Phi_{ww} \right], \cr
& \Phi(0) = 0,  \quad \Phi(c/r) = 1.
}
\right.
\eqno(5.5)
$$
It follows from Lemma 2.1 that $\phi = \Phi$ on $[0, c/r]$.  \qed

We combine the results of Propositions 5.1 and 5.2 in the following theorem.

\thm{5.3} {If $c > rb$, then the maximum probability of reaching the bequest goal is given by
$$
\phi(w) = 
\cases{
{c \over r} \, {(\al_1 - 1)(1 - \al_2) \over \al_1 - \al_2} \left[- \left( {z \over \zo} \right)^{\al_1 - 1} + \left( {z \over \zo} \right)^{\al_2 - 1} \right] z, &if $0 \le w \le b$, \cr \cr
1 - \left( {c \over r} - b \right) {\zb \over p} \left( {{c \over r} - w \over {c \over r} - b} \right)^p, &if $b < w \le {c \over r}$.
}
\eqno(5.6)
$$
in which $\zo$ is given in Proposition $5.1$, and in which $p = {\al_1 \over \al_1 - 1} > 1$. Here, for a given $w \in [0, b]$, $z \in [\zb, \zo]$ uniquely solves $(4.14)$.  When wealth equals $w$, the optimal amount invested in the risky asset is given by
$$
\pi^*(w) = 
\cases{
{\mu - r \over \sig^2} \, {c \over r} \, {(\al_1 - 1)(1 - \al_2) \over \al_1 - \al_2} \left[ \al_1 \left( {z \over \zo} \right)^{\al_1 - 1} - \al_2 \left( {z \over \zo} \right)^{\al_2 - 1} \right], &if $0 \le w < b$, \cr \cr
{\mu - r \over \sig^2} \, {{c \over r} - w \over p - 1}, &if $b < w \le {c \over r}$.
}
\eqno(5.7)
$$}

\rem{5.1} {We find it interesting that the optimal investment strategy when wealth is greater than the bequest goal $b$ is {\it identical} to the corresponding one for minimizing the probability of lifetime ruin, (Young, 2004), which is independent of the ruin level.  Once wealth is greater than the bequest goal $b$, our individual invests like someone who is minimizing the probability of lifetime ruin. \qed}

\rem{5.2} {Browne (1997, Section 3.1) considers a problem related to the one in this section, specifically, maximizing the probability that wealth reaches any $b < {c \over r}$ before $a < b$, for an infinitely lived individual, that is, $\la = 0$.  The game stops as soon as wealth reaches $a$ or $b$.  The optimal amount to invest in the risky asset is ${\mu - r \over \sig^2 (p - 1)} \big |_{\la = 0} \cdot \left( {c \over r} - w  \right) = {2r \over \mu - r} \left( {c \over r} - w \right)$, in which $p = {\al_1 \over \al_1 - 1}$, which is {\it identical} to the optimal investment strategy to minimize the probability of lifetime ruin when $\la = 0$, (Young, 2004). \qed}

For wealth between $0$ and $b$, the optimal investment strategy given in (5.7) is {\it identical} to the one given in (4.15); therefore, many of the properties that we deduced in Section 4.3 for $\pi^*(w)$ when $0 < c \le rb$ hold for $w \in [0, b)$ when $c > rb$.  In particular, as we proved in Proposition 4.4, the optimal investment strategy in (5.7), for wealth less than $b$, is {\it independent} of $b$, a remarkable result.  

For the sake of space, we do not include the analogs of Propositions 4.5, 4.6, and 4.7 here.  Rather, we include two propositions related to the specific case considered in this section, namely, when $c > rb$.  In the first proposition, we show that, as $b$ approaches $0$, then $\phi$ and $\pi^*$ approach $1$ minus the value function and the optimal investment strategy for the problem of minimizing the probability of lifetime ruin, respectively, (Young, 2004).

\prop{5.4} {As $b$ approaches $0$, $\phi$ and $\pi^*$ approach
$$
1 - \left( 1 - {rw \over c} \right)^p,
$$
and
$$
{\mu - r \over \sig^2} \, {{c \over r} - w \over p - 1},
$$
respectively, for $0 < w < c/r$.}

\pf  To prove this proposition, it is enough to show that
$$
\lim_{b \rightarrow 0} \left( {c \over r} - b \right) \zb = p.
\eqno(5.8)
$$
To that end, note that as $b$ approaches $0$, $\zbo$ approaches $1$.  Then, from (5.3), we see that $\zo$ approaches ${rp \over c}$, which is also $\zb$'s limit because $\zbo = \zb/\zo$.  Thus, we have shown (5.8).  \qed

Proposition 5.4 tells us that the solution given in Theorem 5.3 is continuous at $b = 0$.  Also, by comparing Theorems 4.3 and 5.3, we see that $\phi$ and $\pi^*$ are continuous at $c = rb$.  In the second proposition specific to the case for which $c > rb$, we compare $\pi^*(b-)$ with $\pi^*(b+)$.

\prop{5.5} {If $c > rb$, then
$$
\pi^*(b-) = {\mu - r \over \sig^2} \, {c \over r} \, {(\al_1 - 1)(1 - \al_2) \over \al_1 - \al_2} \left( \al_1 \zbo^{\al_1 - 1} - \al_2 \zbo^{\al_2 - 1} \right),
$$
and
$$
\pi^*(b+) = {\mu - r \over \sig^2} \, {{c \over r} - b \over p - 1},
$$
with $\pi^*(b-) > \pi^*(b+)$.}

\pf The expressions for $\pi^*(b-)$ and $\pi^*(b+)$ follow readily from (5.7).  To show that $\pi^*(b-) > \pi^*(b+)$, replace ${c \over r} - b$ in $\pi^*(b+)$ with the left side of (4.7), and simplify to see that the desired inequality is equivalent to $-\al_2 > \al_2$, which is true because $\al_2 < 0$.  \qed

\rem{5.3} {We expect $\pi^*(b-) > \pi^*(b+)$ because for wealth less than $b$, the investor must take on more financial risk to reach her bequest goal.  Once her wealth is greater than $b$, she becomes more conservative and seeks to preserve her wealth while consuming, as in the problem of minimizing the probability of lifetime ruin.  \qed}

\sect{6. Relationship with work of Browne (1997)}

In this paper, we maximize the probability of reaching a specific bequest goal $b > 0$.  Our problem is related to, but different from, the goal-reaching problems of Browne (1997).  First, Browne (1997, Section 3.1) maximizes the probability that wealth reaches $b < c/r$ before it reaches $a < b$.  Browne's game ends when wealth reaches $b$.  By contrast, for the problem we consider, the game continues until the individual dies or until wealth reaches 0.  For further discussion, see Remark 5.2.

Second, Browne (1997, Section 4.2) maximizes the discounted reward of achieving a goal $b \ge c/r$ if $W_ 0 \in [c/r, b]$; if one interprets his discount rate as a hazard rate, then our two problems are {\it mathematically} equivalent.  However, the solution in Browne (1997, Section 4.2) implicitly restricts investment strategies to be such that if $W_0 \in [c/r, b]$, then $W_t \in [c/r, b]$ almost surely, for all $t \ge 0$.  By contrast, in Section 4, we do not restrict our investment strategies in this manner and solve the problem even when $W_0 = w < c/r \le b$.  For further discussion, see Remarks 4.1 and 4.4.  We also point out that the case for which $b < c/r$, which we consider in Section 5, is not considered in Browne (1997), except as discussed in Remark 5.2.

Alternatively, Browne's solution in his Theorem 4.2 implicitly treats $c/r$ as a ruin level.  By contrast, our ruin level is $0$ and thereby allows the individual to invest more aggressively because the game continues if wealth drops below $c/r$.  Therefore, the value function for our problem is strictly greater than the one that Browne presents in his Theorem 4.2; see Proposition 4.9 and Remark 4.4.

As noted in Remark 3.1, our solution and Browne's are identical when $c = 0$.  Otherwise, the results in Sections 3.2, 4, and 5 are new, since the problems considered are, in fact, different.  Browne considers the problem of reaching a goal $b$ before reaching $a < b$, in which $a = c/r$ in his Section 4.2.  We, on the other hand, consider the goal of attaining the bequest $b$ at death (reaching it and then falling below it later does not count) before ruin, which is when wealth hits $0$.  We do this for all levels of $b > 0$.  In the special case for which $b > c/r$ and for which Browne's discount factor equals the individual's hazard rate, reaching $b$ for the first time would be the same problem as having to attain the goal precisely at the time of death; therefore, one would expect that Browne's and our solution to be the same.  However, Browne implicitly assumed $c/r$ to be the ruin level, whereas we take that level to be 0; thus, our solutions differ.

\sect{7. Summary and future work}

We determine the optimal strategy $\pi^*$ for investing in a risky asset in order to maximize the probability $\phi$ of reaching a specific bequest goal $b$.   Here is a summary of our results.

\smallskip

\item{$\bullet$} We obtain closed-form expressions for $\phi$ and $\pi^*$ when the rate of consumption is $0$ and semi-explicit expressions when the rate of consumption is positive.

\item{$\bullet$} For $0 < c \le rb$, we show that the convex Legendre dual of $\phi$ is the value function of an optimal stopping problem.

\item{$\bullet$} For $c > rb$, we show that the convex Legendre dual of $\phi$ is the solution of a time-homogeneous, two-phase Stefan problem. 

\item{$\bullet$} For wealth less than $b$, we show that $\pi^*$ is {\it independent} of $b$.

\item{$\bullet$} For wealth greater than $b$ and less than $c/r$, we show that $\pi^*$ is {\it identical} to the optimal investment strategy when minimizing the probability of lifetime ruin.

\item{$\bullet$} We show that the solution of our problem is continuous at $c = 0$, $c = rb$, and $b = 0$.

\medskip

In future work, we will address a series of problems inspired by this paper.  In particular, we will solve the problem of maximizing the probability of reaching a bequest goal when (1) the market includes life insurance, a financial instrument specifically designed to aid in reaching a bequest goal, (2) consumption is an increasing function of wealth, and (3) life annuities are included in the financial market to cover some or all of consumption.

\bigskip

\centerline{\bf Acknowledgments} \medskip  Research of the first author is supported in part by the National Science Foundation under grant DMS-0955463 and by the Susan M. Smith Professorship of Actuarial Mathematics. Research of the second author is supported in part by the Cecil J. and Ethel M. Nesbitt Professorship of Actuarial Mathematics.

\sect{References}

\noindent \hangindent 20 pt Bayraktar, Erhan, S. David Promislow, and Virginia R. Young (2014a), Purchasing life insurance to reach a bequest goal, {\it Insurance: Mathematics and Economics}, 58: 204-216.

\smallskip \noindent \hangindent 20 pt Bayraktar, Erhan, S. David Promislow, and Virginia R. Young (2014b), Purchasing life insurance to reach a bequest goal: time-dependent case, to appear in {\it North American Actuarial Journal}.

\smallskip \noindent \hangindent 20 pt  Bayraktar, Erhan and Virginia R. Young (2007), Correspondence between lifetime minimum wealth and utility of consumption, {\it Finance and Stochastics}, 11 (2): 213-236.

\smallskip \noindent \hangindent 20 pt Browne, Sid (1997), Survival and growth with a liability: optimal portfolio strategies in continuous time, {\it Mathematics of Operations Research}, 22 (2): 468-493.

\smallskip \noindent \hangindent 20 pt Browne, Sid (1999a), Beating a moving target: optimal portfolio strategies for outperforming a stochastic benchmark, {\it Finance and Stochastics}, 3 (3): 275-294.

\smallskip \noindent \hangindent 20 pt Browne, Sid (1999b), Reaching goals by a deadline: digital options and continuous-time active portfolio management, {\it Advances in Applied Probability}, 31 (2): 551-577.

\smallskip \noindent \hangindent 20 pt Dubins, Lester E. and Leonard J. Savage (1965, 1976), {\it How to Gamble if You Must: Inequalities for Stochastic Processes}, 1965 edition McGraw-Hill, New York. 1976 edition Dover, New York.

\smallskip \noindent \hangindent 20 pt Fasano, Antonio and Mario Primicerio (1997), General free-boundary problems for the heat equation, III, {\it Journal of Mathematical Analysis and Applications}, 59: 1-14.

\smallskip \noindent \hangindent 20 pt Karatzas, Ioannis (1997), Adaptive control of a diffusion to a goal, and a parabolic Monge-Ampere-type equation, {\it Asian Journal of Mathematics}, 1: 295-313.

\smallskip \noindent \hangindent 20 pt Karatzas, Ioannis and Steven E. Shreve (1998), {\it Methods of Mathematical Finance}, New York: Springer-Verlag.

\smallskip \noindent \hangindent 20 pt Kulldorff, Martin (1993), Optimal control of favorable games with a time limit, {\it SIAM Journal on Control and Optimization}, 31 (1): 52-69.

\smallskip \noindent \hangindent 20 pt Merton, Robert C. (1969), Lifetime portfolio selection under uncertainty: the continuous-time case, {\it Review of Economics and Statistics}, 51 (3): 247-257.

\smallskip \noindent \hangindent 20 pt Orey, Steven, Victor C. Pestien, and William D. Sudderth (1987), Reaching zero rapidly, {\it SIAM Journal on Control and Optimization} 25 (5): 1253-1265.

\smallskip \noindent \hangindent 20 pt Peskir, Goran and Albert Shiryaev (2006), {\it Optimal Stopping and Free-Boundary Problems}, Basel: Birkh\"auser Verlag.

\smallskip \noindent \hangindent 20 pt Pestien, Victor C. and William D. Sudderth (1985), Continuous-time red and black: how to control a diffusion to a goal, {\it Mathematics of Operations Research}, 10 (4): 599-611.

\smallskip \noindent \hangindent 20 pt Sudderth, William D. and Ananda Weerasinghe (1989), Controlling a process to a goal in finite time, {\it Mathematics of Operations Research}, 14 (3): 400-409.

\smallskip \noindent \hangindent 20 pt Young, Virginia R. (2004), Optimal investment strategy to minimize the probability of lifetime ruin, {\it North American Actuarial Journal}, 8 (4): 105-126.

 \bye